\documentstyle[12pt]{article}
\input epsf

\newbox\tempboxa
\newdimen\captionboxsubcount \captionboxsubcount=32pt
\newdimen\captionboxsub
\captionboxsub=\hsize \advance\captionboxsub by -\captionboxsubcount
\advance\captionboxsub by -\captionboxsubcount
\catcode`\@=11
\long\def\@makecaption#1#2{
 \vskip 9pt
 \setbox\@tempboxa\hbox{#1: #2}
 \ifdim \wd\@tempboxa >\captionboxsub
\rightskip=\captionboxsubcount \leftskip=\captionboxsubcount
#1: #2
\else \hbox to\hsize{\hfil\box\@tempboxa\hfil}
 \fi}
\catcode`\@=12

\begin{document}
\renewcommand{\thefootnote}{\fnsymbol{footnote}}
\renewcommand{\theequation}{\thesection.\arabic{equation}}
\newcommand{\reseteqnum}{\setcounter{equation}{0}}


\setcounter{footnote}{3}
\renewcommand{\thepage}{}
\begin{titlepage}
\title{
\hfill
\parbox{4cm}{\normalsize KUNS-1270\\HE(TH)~94/09\\hep-ph/9406402}\\
\vspace{1cm}
 QCD $S$ Parameter from Inhomogeneous Bethe-Salpeter Equation
}
\author{
Masayasu Harada\thanks{
Fellow of the Japan Society for the Promotion
of Science for Japanese Junior Scientists.\hfill\break
\indent\quad e-mail
address:{\tt harada@gauge.scphys.kyoto-u.ac.jp }}
{}~and~
Yuhsuke Yoshida\thanks{e-mail
address:{\tt yoshida@gauge.scphys.kyoto-u.ac.jp}}
\\
{\normalsize\em Department of Physics, Kyoto University}\\
{\normalsize\em Kyoto 606-01, Japan}}
\date{\normalsize June, 1994}
\maketitle
\begin{abstract}
We calculate the low-energy parameter $S$ in QCD, which is also known
as $L_{10}$, and the pion decay constant $f_\pi$ using inhomogeneous
Bethe-Salpeter equation in improved ladder approximation.
To extract these quantities we calculate the ``$V-A$'' two-point
function, $\Pi_{VV}(q^2) - \Pi_{AA}(q^2)$, in space-like region.
We obtain $S = 0.43 \sim 0.48$, which is about 30\% larger than
the experimental value.
The calculated $f_\pi$ is well consistent with the result by solving
the homogeneous Bethe-Salpeter equation for pion.
We also evaluate $S$ parameter in $SU(3)$ gauge theory with $N_D$
doublets of fermions in connection with walking technicolor model,
and find that the value of $S/N_D$ hardly depends on $N_D$.
\end{abstract}
\end{titlepage}
\renewcommand{\thepage}{\arabic{page}}
\newpage
\setcounter{footnote}{4}
\setcounter{page}{1}
\section{ Introduction }
\reseteqnum

There is much interest to investigate the low-energy dynamics of QCD.
We can see its property from the low-energy parameters of the
effective Lagrangian such as $L_1$, $L_2$, ..., $L_{10}$, introduced
by Gasser-Leutwyler\cite{GL},
as well as the pion decay constant $f_\pi$.
The parameter $L_{10}$ is related to the $S$ parameter,
which expresses one of the oblique corrections\cite{HT,PT,AB} in
electroweak theory.

These low-energy parameters are calculated by various models.
For example the free quark model gives a half of the
experimental value for QCD $S$ parameter.
It is shown that the parameter $L_{10}$ as well as the other
parameters ($L_1$, $L_2$ and $L_9$) are saturated by the contribution
from the low-lying vector and axial-vector mesons,
$\rho$ and $a_1$.\cite{Ecker-Gasser-Leutwyler-Pich-Derafael}
In Ref.~\cite{Holdom-Terning-Verbeek:PLB245,Holdom},
based on the nonlocal constituent-quark model,
QCD $S$ parameter is calculated using a momentum dependent quark
mass function.
The estimated QCD $S$ parameter well reproduces the
experimental value.
In Ref.~\cite{Donoghue-Holstein} they argue the corrections
to the free quark loop diagram, and
conclude that it is important to include the interaction which forms
bound states as well as corrections to the quark propagator.

In this paper,
we calculate QCD $S$ parameter (i.e., $L_{10}$) using the
inhomogeneous Bethe-Salpeter (BS) equation in the improved ladder
approximation.
The quark mass function is consistently calculated by Schwinger-Dyson
(SD) equation with the same integral kernel.
Our treatment here can include the effects of vector and axial-vector
mesons.
The QCD $S$ parameter is given by the slope of the spin-1 part of
the ``$V-A$'' two-point function, $\Pi_{VV}(q^2)-\Pi_{AA}(q^2)$, at
$q^2=0$.
We obtain the value $S = 0.43 \sim 0.48$.
This is about 30\% larger than the experimental value,
$S = 0.31 \sim 0.38$ \cite{GL},
which comes from the form factors of radiative pion decay
$\pi \rightarrow e\nu\gamma$ and electromagnetic charge radius of
pion.
Noting that from first Weinberg sum rule\cite{Weinberg}
the value of this function at $q^2=0$ gives the pion decay constant
$f_\pi$,
we also calculate the value and show that it is consistent with the
previous result obtained by solving the homogeneous BS equation for
pion.\cite{A-B-K-M-N}
We extract the $\rho$ meson mass and decay constant by three-pole
fitting from the two-point function in the space-like region
($0 \leq -q^2 \leq (1 \mbox{GeV})^2$).

Improved ladder approximation was first used to study the
SD equations\cite{Higashijima,Miransky}, and well succeeded to
describe the property of the chiral symmetry breaking.
The homogeneous BS equation in chiral limit was solved in this
approximation, which represents fundamental properties of
pion.\cite{A-B-K-M-N,JM}
Especially BS equation for pion was extensively studied in QCD and its
generalized model, and the ratio between the pion decay constant
$f_\pi$ and the vacuum expectation value
$\langle\overline{\psi}\psi\rangle$ was calculated in
Ref.~\cite{A-B-K-M-N}.
Moreover, the inhomogeneous BS equation in improved ladder
approximation led to good predictions for low-lying meson
masses.\cite{A-K-M}

We also show the value of $S$ parameter in other dynamical systems.
We evaluate $S$ parameter in $SU(3)$ gauge theory with
$N_D ( = 1, \cdots, 6)$ iso-spin doublets of fermions in connection
with walking technicolor model\cite{H-YBM-AY-AKW}.
Although the evaluation using dynamical mass
function shows that the value of $S/N_D$ is decreased as $N_D$ is
increased\cite{AppelquistTriantaphyllou:PLB278},
we find that the value hardly depends on $N_D$ in the improved ladder
approximation.

This paper is organized as follows.
In section~2, we briefly review the spectral representation of $S$
parameter and $f_\pi$.
We show how to calculate the two-point function $\Pi_{VV}-\Pi_{AA}$
from the inhomogeneous BS amplitude.
Section~3 is devoted to formulations of the inhomogeneous BS
equation.
In section~4 we show the basic tools for the inhomogeneous BS equation
and solve it numerically.
Section~5 is the main part of this paper, where we show the result of
the value of QCD $S$ parameter.
Further we perform three-pole fitting to the
two-point function.
We also investigate the walking coupling case.


\section{ QCD $S$ Parameter and Two-Point Function }
\reseteqnum

In this section, we define the system which we consider in this paper
and summarize the basic ingredients concerning vector and axial-vector
two-point functions for calculating the QCD $S$ parameter.

Supposing that $u$ and $d$ quarks are massless in QCD Lagrangian,
we have the chiral $SU(2)_L \times SU(2)_R$ symmetry.
As is well known, this symmetry is spontaneously broken down to its
subgroup $SU(2)_V$, and massless pions appear.
Chiral Lagrangian well represents the symmetrical aspects of the
interaction among pions and external currents which couple to photon,
$W$ and $Z$ bosons.
There are several low-energy parameters in the effective chiral
Lagrangian and these parameters are to be determined from the dynamics
of QCD.

For calculating QCD $S$ parameter we consider the vector and
axial-vector current defined by
\begin{equation}
\left\{
\begin{array}{ll}
\displaystyle V^a_\mu(x) =
   \overline\psi(x)\frac{\tau^a}{2}\gamma_\mu \psi(x)~, \medskip\\
\displaystyle A^a_\mu(x) =
   \overline\psi(x)\frac{\tau^a}{2}\gamma_\mu\gamma_5 \psi(x)~,
\end{array}\right.\qquad\qquad
\psi = \left(\begin{array}{c}
u \\ d \end{array}\right)~,\\
\end{equation}
where $\tau^a$ ($a = 1,2,3$) is Pauli matrix.
Then we define the two-point function $\Pi_{JJ}$ ($J = V, A$)
\begin{eqnarray}
\delta^{ab}\Pi_{JJ}(q^2) &\equiv&
   \epsilon^\mu\epsilon^\nu i \int d^4x \: e^{iqx}
   \left\langle 0\right|{\rm T} J^a_\mu(x) J^b_\nu(0) \left|
0\right\rangle~,\nonumber\\
J^a_\mu(x) &\equiv&
   V^a_\mu(x)\,,\: A^a_\mu(x)~,\label{eq. def pi}
\end{eqnarray}
where $a$, $b$ are iso-spin indices, $\epsilon_\mu$ is the polarization
vector defined by $\epsilon\cdot q = 0$, $\epsilon\cdot\epsilon = -1$.

The QCD $S$ parameter and the pion decay constant are given by the
two-point function of ``$V-A$'' type:
\begin{eqnarray}
S      &=&  4\pi \left.\frac{d}{dq^2}
   \left[ \Pi_{VV}(q^2) - \Pi_{AA}(q^2) \right]
   \right\vert_{q^2 = 0}~,\label{eq. S from v-a}\\
f_\pi^2 &=& \Pi_{VV}(0) - \Pi_{AA}(0) ~.\label{eq. fpi from v-a}
\end{eqnarray}
QCD $S$ parameter is related to the Gasser-Leutwyler parameter
$L_{10}$ as $S = -16\pi L_{10}$.

Using spectral representation of vector and axial-vector currents,
eqs. (\ref{eq. S from v-a}) and (\ref{eq. fpi from v-a}) are rewritten
into
\begin{eqnarray}
S &=&
   4\pi\int_0^\infty\frac{ds}{s^2}\:
\left[ \rho_{_V}(s) - \rho_{_A}(s) \right]~,\label{eq. pip0}\\
f_\pi^2 &=&
   \int_0^\infty \frac{ds}{s}\: \left[ \rho_{_V}(s) - \rho_{_A}(s)
\right]~,\label{eq. pi0}
\end{eqnarray}
where $\rho_{_V}(s)$ and $\rho_{_A}(s)$ are the spin-1 parts of the vector
and axial-vector spectral functions, respectively.
Equations (\ref{eq. pip0}) and (\ref{eq. pi0}) are
referred to as the Das-Mathur-Okubo sum rule\cite{D-M-O}
and the first Weinberg sum rule\cite{Weinberg}, respectively.
We can easily understand the above equations by the following way:
The commutator of the conserved current $J^a_\mu(x)$ is decomposed
into
\begin{eqnarray}
\left\langle 0\right|\left[ J^a_\mu(x) ,
J^b_\nu(0) \right] \left| 0\right\rangle &=&
   -\delta^{ab} \int_0^\infty \frac{ds}{s}\:
   \rho_{_J}(s) ( s g_{\mu\nu}+ \partial_\mu \partial_\nu )
	\Delta(x;s) \nonumber\\
   &&- \delta^{ab} \; \rho^{(0)}\; \partial_\mu \partial_\nu \Delta(x;0)~,\\
   \Delta(x;s) &=& \int \frac{d^4q}{(2\pi)^3}
   e^{-iqx}\epsilon(q_0)\delta(s-q^2)~, \nonumber
\end{eqnarray}
where $\rho^{(0)}$ denotes a contribution from massless scalar
particle.
Noting that massless pion couples to the axial-vector current
$A^a_\mu$ while no massless particle couples to the vector current
$V^a_\mu$, we find $\rho^{(0)}_{_A} = f_\pi^2$ and
$\rho^{(0)}_{_V} = 0$.
The ``$V-A$'' two-point function can be expressed as
\begin{eqnarray}
\lefteqn{
i \int d^4x \: e^{iqx} \Bigl\langle 0\Bigr\vert{\rm T}\left[ V^a_\mu(x)
V^b_\nu(0) -
   A^a_\mu(x) A^b_\nu(0) \right]\Bigl\vert 0\Bigr\rangle }\nonumber\\
&=& - \delta^{ab} \int_0^\infty \frac{ds}{s}\: \frac{\rho_V(s) -
\rho_A(s)}{s - q^2 - i\epsilon}
	( s g_{\mu\nu}- q_\mu q_\nu )
 + \delta^{ab} \; \frac{q_\mu q_\nu}{q^2}\,f_\pi^2~.
\label{eq. spectral repr.}
\end{eqnarray}
Thus, we obtain eqs.~(\ref{eq. pip0}) and (\ref{eq. pi0}) using
eqs.~(\ref{eq. S from v-a}) and (\ref{eq. fpi from v-a}).

Let us consider the following three-point vertex function
\begin{eqnarray}
\lefteqn{ \delta_i^j~\left(\frac{\tau^a}{2}\right)_{\!\!\!f}^{f'}
\int \frac{d^4p}{(2\pi)^4}~e^{-ipr}~
\chi_{\alpha\beta}(p;q,\epsilon) ~~ = }\nonumber\\
&&~~~~~~~~ \epsilon^\mu~\int d^4x~e^{iqx}~
\Bigl\langle{0}\Bigr| {\rm T}\,\psi_{\alpha if}(r/2)\,
   \overline\psi_\beta^{jf'}\!\!(-r/2)\,J^a_\mu(x)
   \Bigl|{0}\Bigr\rangle~,
\end{eqnarray}
where $\chi$ ($=\!\chi_{_V},\chi_{_A}$) is bi-spinor, which we call
inhomogeneous BS amplitude.
Here ($f$, $f'$, $\cdots$), ($i$, $j$, $\cdots$)
and ($\alpha$, $\beta$, $\cdots$) denote flavor, color and spinor
indices, respectively.
This inhomogeneous BS amplitude has definite spin, parity and charge
conjugation, i.e., $J^{PC} = 1^{--}$ for vector case and $1^{++}$ for
axial-vector case.

Closing the fermion legs of the three-point function,
we find that the two-point function is expressed in terms of the
inhomogeneous BS amplitude $\chi$:
\begin{eqnarray}
\Pi_{JJ}(q^2) &=&
   -\frac{1}{3}\sum_\epsilon \int\frac{d^4p}{i(2\pi)^4} \;
   \frac{N_c}{2}~{\rm tr}\Bigl[ (\epsilon \cdot G)\,
\chi(p;q,\epsilon)\Bigr]~,\nonumber\\
G_\mu &=& \left\{ \begin{array}{ll}
   \gamma_\mu         & \mbox{ for vector vertex }\\
   \gamma_\mu\gamma_5 & \mbox{ for axial-vector vertex }
   \end{array}\right.~,
\end{eqnarray}
where $N_c \!\equiv\! 3$ is the number of $SU(3)_c$ color and we average
over the polarizations for convenience;
$\Pi_{JJ}(q^2)$ in eq.~(\ref{eq. def pi}) does not depend on the
polarization $\epsilon_\mu$.
Although the vector or axial-vector two-point function itself is
logarithmically divergent,
the chiral symmetry guarantees the finiteness of the
quantity $\Pi_{VV}(q^2) - \Pi_{AA}(q^2)$.
The divergences appearing in each two-point function cancel
because the structures of the divergences are exactly the
same.\cite{Inami-Lim-Yamada}

The quantities we should first calculate are vector and
axial-vector inhomogeneous BS amplitudes $\chi_{_V}$ and $\chi_{_A}$,
which are finite by current conservation.
Then, we perform the momentum integration after taking the difference
to obtain the ``$V-A$'' two-point function, i.e.,
\begin{equation}
\Pi_{VV}(q^2) - \Pi_{AA}(q^2) =
   -\frac{1}{3}\sum_\epsilon \int\frac{d^4p}{i(2\pi)^4} \;
   \frac{N_c}{2}~
   {\rm tr}\Bigl[\; { \epsilon\kern-5.2pt\mbox{\it/} \chi_{_V}(p;q,\epsilon) -
\epsilon\kern-5.2pt\mbox{\it/}\gamma_5 \chi_{_A}(p;q,\epsilon) }\; \Bigr]~.
\label{eq. two-point function}
\end{equation}
This integration converges as we mentioned above.
In this paper, we work with the space-like total momentum $q_\mu$
($q_E^2 \equiv -q^2 > 0$),
then we do not encounter the singularities which come from meson poles
in time-like region.


\section{ Inhomogeneous Bethe-Salpeter Equation }
\reseteqnum

In this section, we discuss the basic formulations for solving the
vector and axial-vector inhomogeneous BS equations.
SD equation is solved by  the same BS kernel.
We give the component form of the inhomogeneous BS equation.
The inhomogeneous BS equations are solved in the space-like region for
the total momentum $q_\mu$, $q_E^2 > 0$.

\begin{figure}[bhtp]
\begin{center}
\ \epsfbox{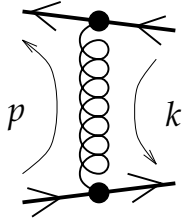}
\vspace{-5pt}
\caption[]{The Feynman diagram of the BS kernel $K(p,k)$
in eq.~(\ref{eq. BS kernel}) which we use in our calculation.
The helix denotes the gluon propagator intermediating between two
quarks.}
\label{fig. BS kernel}
\end{center}
\end{figure}

In the non-perturbative treatment of QCD by BS approach,
the most important quantity is the BS kernel 
which expresses the QCD interaction by the gluon.
In the improved ladder approximation with Landau gauge,
the BS kernel $K$ is defined by
[the momentum assignment is chosen as shown in
Fig.~\ref{fig. BS kernel}]
\begin{equation}
K(p,k) = C_2\, g^2(p,k)
   \frac{1}{-(p-k)^2}\left( g_{\mu\nu} - \frac{(p-k)_\mu (p-k)_\nu}{(p-k)^2}
\right)
   \gamma^\mu \otimes \gamma^\nu~,\label{eq. BS kernel}
\end{equation}
where $C_2 \equiv (N_c^2 - 1)/(2N_c)$ is the
second Casimir of $SU(3)_c$ color fundamental representation
and we use the tensor product notation\cite{K-M-Y}
\begin{equation}
( A \otimes B ) \chi \equiv A \chi B~.
\end{equation}
In the BS kernel (\ref{eq. BS kernel}) we adopt the
Higashijima-Miransky approximation\cite{Higashijima,Miransky}
for the running coupling, $g^2(p,k) = g^2(\max(-p^2,-k^2))$.
This running coupling allows us to include the property of
asymptotic freedom of QCD.
Using this running coupling the chiral symmetry is always
spontaneously broken.
The detailed structure of the running coupling is given in
section~\ref{Running Coupling}.

For solving the inhomogeneous BS equation, we need the full quark
propagator $S_F(p)$, which is given by solving the SD equation
with the {\it same} BS kernel for preserving the chiral symmetry.
The wave function renormalization factor of quark propagator is one
in Landau gauge in the Higashijima-Miransky approximation.
Then the quark mass function $\Sigma(p)$ is given by
\begin{equation}
i \Sigma(p) = K * S_F(p)~,\label{eq. SD}
\end{equation}
where
\begin{equation}
S_F(p) = \frac{i}{p\kern-6.0pt\mbox{\it/} - \Sigma(p)}~.
\end{equation}
The operator ``$*$'' acting on the BS kernel $K$ and a bi-spinor
$\Psi$ denotes momentum integration:
\begin{equation}
K * \Psi(p) \equiv \int \frac{d^4k}{i(2\pi)^4} \; K(p,k) \; \Psi(k)~.
\end{equation}

\begin{figure}[hbtp]
\begin{center}
\ \epsfbox{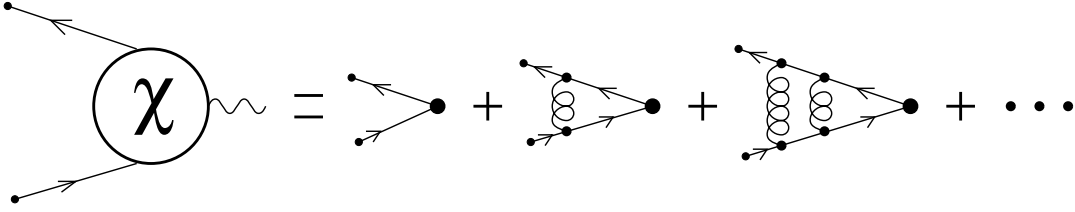}
\caption[]{The expansion of vector or axial-vector
   inhomogeneous BS amplitude given in eq.~(\ref{eq. series}).}
\label{fig. series}
\end{center}
\end{figure}

Now, the inhomogeneous BS equation for $\chi(p;q,\epsilon)$ is
\begin{equation}
( T - K * ) \; \chi = (\epsilon\cdot G)~,
\qquad(~G_\mu = \gamma_\mu, \gamma_\mu\gamma_5~)\label{eq. ibs}
\end{equation}
where
\begin{equation}
T = T(p;q) \equiv S_F^{-1}(p+\frac{q}{2})
   \otimes S_F^{-1}(p-\frac{q}{2}) ~. \\
\end{equation}
The formal solution is given by
\begin{equation}
\chi = \frac{1}{ T - K * } \; (\epsilon\cdot G)~.
\end{equation}
This solution can be reinterpreted by expanding it into the power
series of the BS kernel as (see Fig.\ref{fig. series})
\begin{equation}
\chi = T^{-1}(\epsilon\cdot G) + T^{-1}K*T^{-1}(\epsilon\cdot G)
   + T^{-1}\left(K*T^{-1}\right)^2(\epsilon\cdot G) + \cdots ~.\label{eq.
series}
\end{equation}

We can expand the inhomogeneous BS amplitude
$\chi = \chi_{{}_V},\,\chi_{{}_A}$ into eight invariant amplitudes:
\begin{equation}
\chi_{{}_J}(p;q,\epsilon) =
\displaystyle\sum_{i = 1}^8 \Gamma^{(J)}_i(p;q,\epsilon)~
   \chi^i_{{}_J}(p;q)~,\qquad (~J = V, A~)
\end{equation}
where $\chi^i_{{}_J} \: (i = 1,\cdots, 8)$ is scalar quantity.
$\Gamma^{(J)}_i$ is the vector or
axial-vector base defined by
\begin{equation}
\begin{array}{llll}
\Gamma^{(V)}_1 = \epsilon\kern-5.2pt\mbox{\it/} , &
\Gamma^{(V)}_2 = \frac{1}{2}[\epsilon\kern-5.2pt\mbox{\it/}
,p\kern-6.0pt\mbox{\it/}]
   (p\cdot\widehat{q}) , &
\Gamma^{(V)}_3 = \frac{1}{2}[\epsilon\kern-5.2pt\mbox{\it/}
,\widehat{q}\kern-5.8pt\mbox{\it/}] , &
\Gamma^{(V)}_4 = \frac{1}{3!}[\epsilon\kern-5.2pt\mbox{\it/}
,p\kern-6.0pt\mbox{\it/},\widehat{q}\kern-5.8pt\mbox{\it/}] ,\bigskip\\
\Gamma^{(V)}_5 = (\epsilon\cdot p) , &
\Gamma^{(V)}_6 = p\kern-6.0pt\mbox{\it/}(\epsilon\cdot p) , &
\Gamma^{(V)}_7 =
\widehat{q}\kern-5.8pt\mbox{\it/}(p\cdot\widehat{q})(\epsilon\cdot p) , &
\Gamma^{(V)}_8 = \frac{1}{2}[p\kern-6.0pt\mbox{\it/},
\widehat{q}\kern-5.8pt\mbox{\it/}](\epsilon\cdot p) , \bigskip
\end{array} \label{eq. v base}
\end{equation}
and
\begin{equation}
\Gamma^{(A)}_i = \Gamma^{(V)}_i \gamma_5~, \label{eq. a base}
\end{equation}
where $\widehat{q}_\mu = q_\mu/\sqrt{q_E^2}$ and $[a,b,c] \equiv
a[b,c] + b[c,a] + c[a,b]$.
We note that the dependence on the polarization vector $\epsilon_\mu$
is isolated in the base $\Gamma_i^{(J)}(p;q,\epsilon)$.

To solve the integral equation (\ref{eq. ibs}) numerically,
we perform the Wick rotation on the momentum $k$ integration
and analytic continuation on the relative momentum $p$ as usual.
We introduce the scalar variables $u$, $x$, $v$ and $y$ as
\begin{equation}
\begin{array}{cc}
p\cdot \widehat{q} = -u , & p^2 = -u^2-x^2 , \\
k\cdot \widehat{q} = -v , & k^2 = -v^2-y^2   ~.
\end{array}
\end{equation}
Multiplying eq.~(\ref{eq. ibs}) by the Dirac conjugate base
$\overline\Gamma_i$,
taking the trace and summing over the polarization,
we convert it into the component form%
\footnote{We evaluate the matrix elements of $T_{ij}(u,x)$ and
$K_{ij}(u,x;v,y)$ by an algebraic calculation program.}
\begin{equation}
\sum_j~( T_{ij} - K_{ij} \star )\, \chi^j = I_i~,\label{eq. ibs matrix}
\end{equation}
where
\begin{eqnarray}
I_i(u,x) &=& \frac{1}{4}\sum_\epsilon {\rm tr}\left[\;
   { \overline\Gamma_i(p;\widehat{q},\epsilon)\, (\epsilon\cdot G) }
   \;\right]~,\nonumber\\
T_{ij}(u,x) &=& \frac{1}{4}\sum_\epsilon {\rm tr}\left[\;
{ \overline\Gamma_i(p;\widehat{q},\epsilon)\,
   T(p;q)\, \Gamma_j(p;\widehat{q},\epsilon)  }
   \;\right]~,\nonumber\\
K_{ij}(u,x;v,y) &=& 2\pi\int_{-1}^1d\cos\theta\;
   \frac{1}{4}\sum_\epsilon {\rm tr}\left[\; {
\overline\Gamma_i(p;\widehat{q},\epsilon)\,
   K(p,k) \,\Gamma_j(k;\widehat{q},\epsilon)} \;\right]~.\qquad
\end{eqnarray}
Here $\theta$ is the angle between the 3-vector part of $p$ and $k$;
$\cos\theta \equiv \mbox{\boldmath $p$} \cdot \mbox{\boldmath $k$}
/|\mbox{\boldmath $p$}||\mbox{\boldmath $k$}|$.
The operation $\star$ denotes $u$ and $x$ integrations, i.e.,
$2\pi\int_{-1}^1d\cos\theta \;\star \equiv *$.
We should note that the Dirac conjugation is taken to be
\begin{equation}
\overline \chi(p;q,\epsilon)
   \equiv
   \gamma_0 \chi(p^*;q^*,\epsilon)^\dagger \gamma_0~.
\end{equation}
The complex conjugate on $p$ and $q$ should be taken to preserve
Feynman causality of inhomogeneous BS amplitudes when those momenta
become complex by analytic continuation.\cite{K-M-Y}
Our choice of the bases (\ref{eq. v base}) and (\ref{eq. a base})
leads to the fact that
the matrix $K_{ij}$ is independent of $q_E^2$.
[The $q_E^2$ dependence of $\chi^j$ comes from $T_{ij}$ only.]

It should be noticed that the invariant amplitude $\chi^i$ is even
function of ($p\cdot\widehat{q}$).
Namely it is regarded as an even function of $u$ with an arbitrary
constant $q_E^2$:
\begin{equation}
\chi^i(u,x) = \chi^i(-u,x)~.\label{eq. chi even}
\end{equation}
This is the result of the charge conjugation property
\begin{equation}
\begin{array}{c}
C \chi(-p;q,\epsilon)^T C^{-1} = -\chi(p;q,\epsilon)~,\smallskip\\
C \Gamma_i(-p;q,\epsilon)^T C^{-1} = -\Gamma_i(p;q,\epsilon)~,
\end{array}
\end{equation}
where $C =  i\gamma_0\gamma_2$ is charge conjugation matrix.
Similarly from this property,
one can easily check that $T$ and $K$ are real definite and symmetric
matrices:
\begin{eqnarray}
T_{ij}(u,x) &=& T_{ji}(u,x)~,\nonumber\\
K_{ij}(u,x;v,y) &=& K_{ji}(v,y;u,x)~.
\end{eqnarray}
This is an important property, from which we find that the
inhomogeneous BS amplitudes are real definite, and thus the
two-point function is real definite.


\section{ Numerical Calculations }
\reseteqnum

In this section we give the detailed form of the running coupling.
We calculate the quark mass function and the inhomogeneous BS
amplitudes.
In the following, all the dimensionful parameters are rescaled by
$\Lambda_{\rm QCD}$, otherwise stated.

\subsection{ Running Coupling \label{Running Coupling} }

Running coupling $g^2(\mu^2)$ can be well approximated by the result
from one-loop $\beta$-function in high energy region,
so we basically use it in the BS kernel (\ref{eq. BS kernel}).
However the one-loop running coupling blows up at
$\mu = \Lambda_{\rm QCD}$
and we have no idea about the functional form in low energy region.
One prescription is to adopt the Higashijima
approximation\cite{Higashijima} in which
$g^2(\mu^2)$ is constant below some scale and is one-loop running
coupling above that scale.
It is important that the running coupling and its derivative
are continuous with respect to $\ln\mu^2$, otherwise the derivative of
the mass function is discontinuous.\footnote[2]
{We easily find this point if we convert the SD equation into
differential equation.}
We achieve this continuities by interpolating between one-loop running
coupling and a fixed value with the second order polynomial of
$\ln\mu^2$.
According to Ref.~\cite{A-B-K-M-N},
we adopt the following functional form of the running coupling:
\begin{equation}
\alpha(\mu^2) \equiv \frac{g^2(\mu^2)}{4\pi} =
   \alpha_0 \times \left\{\begin{array}{ll}
\displaystyle \frac{1}{t} & \mbox{ if $t_F < t$ } \smallskip\\
\displaystyle \frac{1}{t_F} + \frac{(t_F - t_C)^2
   - (t - t_C)^2}{2t_F^2(t_F - t_C)} &\smallskip
   \mbox{ if $ t_C < t < t_F$ } \\
\displaystyle \frac{1}{t_F} + \frac{(t_F - t_C)^2}{2t_F^2(t_F - t_C)} &
   \mbox{ if $ t < t_C$ } \smallskip
   \end{array}\right.~,\label{eq. alpha}
\end{equation}
where $t = \ln \mu^2$ and $\alpha_0 = 12\pi/(11N_c - 2N_f)$ with $N_f$
being the number of flavors.
As seen in Fig.~\ref{fig. support}, the support of the
``$V-A$'' two-point function lies below in the threshold of $c$ quark,
so we take $N_f = 3$ and $\alpha_0 = 4\pi/9$.
We take the same parameter choice as in Ref.~\cite{A-K-M}, i.e.,
$t_F=0.5$ and $t_C=-2.0$.
It is observed at least that the specific choice of the parameter
$t_F$ does not affect the ``physical'' quantity
$\langle\overline\psi\psi\rangle$.\cite{A-B-K-M-N}
We will check the dependence on the infrared cutoff $t_F$ of the QCD
$S$ parameter in the section~\ref{Results}.

\subsection{ Mass Function }

Before solving the inhomogeneous BS equation, we have to calculate
the mass function of quarks.
The SD equation (\ref{eq. SD}) reads
\begin{equation}
\Sigma(x) = \frac{3C_2}{4\pi}\int_0^\infty dy~
   \frac{\alpha(\max(x,y))}{\max(x,y)}~
   \frac{y\Sigma(y)}{y + \Sigma^2(y)}\label{eq. lattice SD}
\end{equation}
in the Higashijima-Miransky approximation,
where $x \!=\! -p^2$ and $y \!=\! -k^2$.
This integral equation is solved by the following way:
First, we discretize the equation fine enough.
Second, starting from the functional form
$\Sigma(x) = \mbox{constant} \not= 0$,
we iteratively update the mass function according to
eq.~(\ref{eq. lattice SD}) itself
until the functional form converges.

When we solve the inhomogeneous BS equation,
we need the mass function at the momenta $p \pm q/2$.
To obtain the mass function $\Sigma(-(p \pm q/2)^2)$,
we substitute the above convergent mass function $\Sigma(y)$ into RHS
of eq.~(\ref{eq. lattice SD}) after putting $x = -(p\pm q/2)^2$, and
carry out the $y$ integration.
We note that we can independently choose the lattice points for
solving the SD and inhomogeneous BS equations.

\subsection{ Inhomogeneous BS Amplitude }

Let us consider the inhomogeneous BS equation (\ref{eq. ibs matrix}).
We discretize it and solve the resulting linear equation numerically.
Here, we should note that $\chi^j(v,y)$ is even function of $v$.
Then we restrict the integral region over variable $v$ to be
positive, $v > 0$, after replacing the kernel as
\begin{equation}
K_{ij}(u,x;v,y) \rightarrow K_{ij}(u,x;v,y) + K_{ij}(u,x;-v,y)~.
\end{equation}

The fundamental variables used to solve the inhomogeneous BS equation
(\ref{eq. ibs matrix}) are $U$ and $X$ defined by
\begin{equation}
\begin{array}{ll}
   u = \exp U~, & x = \exp X
\end{array}~.\label{eq. UX}
\end{equation}
As for the strong interaction, it is important to take into account
interactions around $\Lambda_{\rm QCD}$ scale rather than that in the
high energy scale.
The above choice (\ref{eq. UX}) is suitable for our calculation.
We discretize the variables $U$ and $X$ at $N_{BS} = 22$ points evenly
spaced in the intervals
\begin{equation}
\begin{array}{cl}
   U \in \left[\;\lambda_U\,,\,\Lambda_U\;\right]
   \hspace{-7pt}&= \left[\;-5.5\,,\,2.5\;\right]~, \vspace{12pt}\\
   X \in \left[\;\lambda_X\,,\,\Lambda_X\;\right]
   \hspace{-7pt}&= \left[\;-2.5\,,\,2.5\;\right]~.
\end{array}\label{eq. cutoffs}
\end{equation}

In numerical integration, to avoid integrable logarithmic singularity
at $(u,x) = (v,y)$,
we take four-point average\cite{A-K-M} of the BS kernel
$K_{ij}(u,x;v,y)$ as
\begin{eqnarray}
K_{ij}(u,x;v,y) &\rightarrow& \frac{1}{4}\left[\;
   K_{ij}(u,x;v_+,y_+)\ +\ K_{ij}(u,x;v_+,y_-)\ + \right.\nonumber\\
&&~~~~\left. K_{ij}(u,x;v_-,y_+)\ +\ K_{ij}(u,x;v_-,y_-)
   \;\right]~,
\end{eqnarray}
where
\begin{equation}
\begin{array}{ll}
\displaystyle v_\pm = \exp(V \pm \frac{1}{4}D_U)~, &
\displaystyle D_U = (\Lambda_U - \lambda_U)/(N_{BS} - 1)~,\vspace{12pt}\\
\displaystyle y_\pm = \exp(Y \pm \frac{1}{4}D_X)~, &
\displaystyle D_X = (\Lambda_X - \lambda_X)/(N_{BS} - 1)~.
\end{array}
\end{equation}

Now, we solve the inhomogeneous BS equations for the vector and the
axial-vector currents separately,
so that we obtain the amplitudes $\chi_{{}_V}(u,x)$ and
$\chi_{{}_A}(u,x)$.
We use FORTRAN subroutine package for these numerical calculations.

\section{ Results }\label{Results}
\reseteqnum

In this section, first we calculate the spin-1 part of the two-point
function, $\Pi_{VV}(q^2)-\Pi_{AA}(q^2)$,
then extract the QCD $S$ parameter and the pion decay constant
$f_\pi$.

\subsection{ QCD $S$ Parameter }

After obtaining the vector and axial-vector inhomogeneous BS
amplitudes, $\chi_{_V}(u,x)$ and $\chi_{_A}(u,x)$, numerically,
we calculate the ``$V-A$'' two-point function using
eq.~(\ref{eq. two-point function}).

The QCD $S$ parameter and the pion decay constant are calculated from
the formulae (\ref{eq. S from v-a}) and (\ref{eq. fpi from v-a}).
Using the numerical differentiation formula,
we extract $S$ from four data points of $\Pi_{VV}-\Pi_{AA}$ at
$q^2_E \equiv -q^2 = 0.0, 0.2, 0.4, 0.6$.
In this choice of the interval of $q^2_E$, $h = 0.2$,
the error of numerical differentiation is estimated
as $O(h^3)\sim 0.8\%$.%
\footnote[5]{
   It is natural to estimate the numerical error with the
   dimensionless quantities scaled by $\Lambda_{\rm QCD}$.
}
On the other hand, the fluctuation $\delta$ of the value of
$\Pi_{VV}-\Pi_{AA}$, which comes from the dependence on the lattice size,
causes the error $O(\delta/h)$ of $S$.
As we will show below, choosing large lattice size allows us to make
the fluctuation $\delta$ within $1\%$.
Then the error of $S$ is expected as $O(\delta/h)\sim 5\%$.
When we take larger interval $h$, the error of numerical
differentiation becomes large.
On the other hand, the fluctuation of $S$, $O(\delta/h)$, is enhanced
by taking the smaller interval $h$.
We have checked the validity of the differentiation for several
choices of $h$ and for various numerical differentiation formulae.

In what follows we investigate all the dependences on the parameters,
i.e., the momentum cutoff (\ref{eq. cutoffs}), the lattice size
$N_{BS}$ and the infrared cutoff $t_F$ of the running coupling.

\begin{figure}[hbtp]
\begin{center}
\ \epsfbox{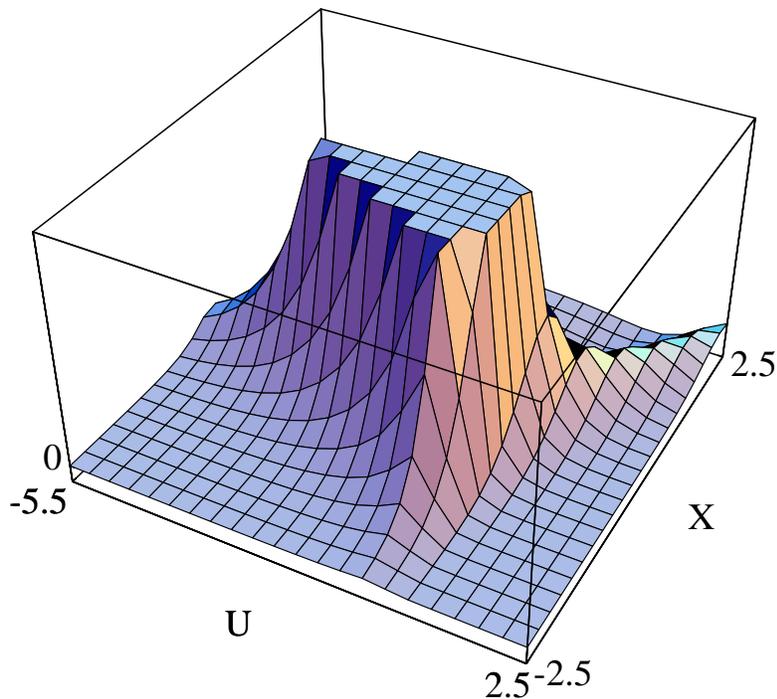}
\vspace{-5pt}
\caption[]{The support of ``$V-A$'' two-point function.
The upper $9/10$ of the figure is cripped.
}
\label{fig. support}
\end{center}
\end{figure}

First, we check the dependence on the infrared and ultraviolet cutoff
in eq.~(\ref{eq. cutoffs}).
The typical example of the support of the ``$V-A$'' two-point function
is shown in Fig.~\ref{fig. support} with $q^2 = 0$, $t_F = 0.5$ and
$N_{BS} = 22$.
The choice (\ref{eq. cutoffs}) covers the dominant support very well.
Further, the position of the support does not change if we vary the
values of $q^2$, $t_F$ and $N_{BS}$.
There is a slight rise in the high energy region, which originates
in the numerical error of the cancellation of divergences between
$\Pi_{VV}$ and $\Pi_{AA}$.
This error, if any, affects the two-point function with 1\% at most.

Second, we check the dependence on the lattice size $N_{BS}$.
We show the values of the QCD $S$ parameter and the pion decay
constant $f_\pi$ for several values of $N_{BS}$
in Table~\ref{tab. NBS dependence}.
We also show the values of $\Lambda_{\rm QCD}$ which are fixed by
imposing $f_\pi = 93$ MeV.
Even in $N_{BS} = 14$ the fluctuation $\delta$ of $f_\pi^2$
is within 1\% and the fluctuation of $S$ is several percents.
So, it is enough to take $N_{BS} = 22$.
\begin{table}[hbtp]
\begin{center}
\begin{tabular}[t]{|c||c|c|c|c|c|c|c|}
\hline
$N_{BS}$
        & 10      & 12     & 14     & 16
   & 18      & 20     & 22      \\\hline\hline
$S$
        & 0.500   & 0.388  & 0.452  & 0.481
   & 0.455   & 0.474  & 0.464\\ \hline

$(f_\pi/\Lambda_{\rm QCD})^2 \times 100$
        & 3.71    & 3.76      & 3.95   & 4.04
   & 4.04    & 4.06   & 4.05    \\ \hline
$\Lambda_{\rm QCD}$ [MeV]
        & 483     & 479    & 468    & 462
   & 463     & 461    & 462     \\ \hline
\end{tabular}
\caption[]{
$N_{BS}$ dependence of the value of the QCD $S$ parameter and
$f_\pi/\Lambda_{\rm QCD}$.
We also show the values of $\Lambda_{\rm QCD}$ calculated by imposing
$f_\pi = 93$ MeV.
We take the parameter choice $t_F = 0.5$.
}\label{tab. NBS dependence}
\end{center}
\end{table}

Third, we check the dependence on the infrared cutoff $t_F$ taking
$N_{BS} = 22$.
We show the values of the QCD $S$ parameter and the pion decay
constant $f_\pi$ with several values of $t_F$
in Table \ref{tab. tF dependence}.
The variations of $S$ are within 10\%.
We also calculate the scale $\Lambda_{\rm QCD}$ by imposing
$f_\pi = 93$~MeV, and the results are shown in
Table~\ref{tab. tF dependence}.
The values of $\Lambda_{\rm QCD}$ are consistent with the
results in Ref.~\cite{A-B-K-M-N} which are calculated from the
homogeneous BS amplitude of pion with the same BS kernel as ours.
For example, their typical value is $\Lambda_{\rm QCD} = 484$~MeV with
$t_F = 0.5$.
[Our $t_F$ corresponds to $t_{IF}$ in Ref.~\cite{A-B-K-M-N} as
$t_F = 1+t_{IF}$.]
\begin{table}[hbtp]
\center{
\begin{tabular}[t]{|c||c|c|c|c|c|}
\hline
$t_F$
   & 0.3    & 0.5    & 0.7    & 0.9    & 1.1    \\\hline\hline
$S$
   & 0.432 & 0.464   & 0.478  & 0.481  & 0.470  \\ \hline
$(f_\pi/\Lambda_{\rm QCD})^2 \times 100$
   & 3.43  & 4.05    & 4.10   & 3.75   & 3.12   \\ \hline
$\Lambda_{\rm QCD}$ [MeV]
   & 502   & 462     & 459    & 481    & 526    \\ \hline
\end{tabular}}
\caption[]{$t_F$ dependence of the value of the QCD $S$ parameter and
$f_\pi/\Lambda_{\rm QCD}$.
We also show the values of $\Lambda_{\rm QCD}$ calculated by imposing
$f_\pi = 93$ MeV.
We fix the lattice size as $N_{BS} = 22$.
}\label{tab. tF dependence}
\end{table}

\begin{figure}[hbtp]
\begin{center}
\ \epsfbox{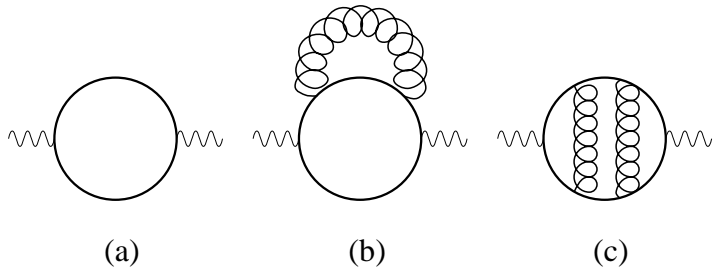}
\vspace{-5pt}
\caption[]{
The schematic view of the Feynman diagrams contributing to the
``$V-A$'' two-point function,
$\Pi(q_E^2) \equiv \Pi_{VV}(q_E^2) - \Pi_{AA}(q_E^2)$.
(a) is quark one-loop diagram,
(b) expresses inclusion of momentum dependent mass function,
(c) represents binding forces to form quark-antiquark bound state.
}\label{fig. loop}
\end{center}
\end{figure}

We show the value of the QCD $S$ parameter which are the main results
of this paper:
\begin{equation}
S = 0.43 \sim 0.48~.\label{eq. S}
\end{equation}
Let us consider what effects are included by our approach.
The lowest contribution to QCD $S$ parameter is given by
one-loop quark diagram shown in Fig.~\ref{fig. loop}(a).
There are two classes of the higher order corrections to this diagram,
i.e., corrections to the quark propagator and binding forces to
form $q\overline q$ bound state.
[Effects of gluon condensation\cite{GLUON} are not considered here.]
These are schematically expressed by the diagrams in
Figs.~\ref{fig. loop}(b) and \ref{fig. loop}(c).
The improved ladder approximation includes these two
corrections simultaneously.

\begin{table}[hbtp]
\begin{center}
QCD $S$ parameters\\
\begin{tabular}[t]{cccccc}
\hline\hline
\raisebox{-8pt}{\ }\raisebox{13pt}{\ }
Our value & GL & VMD & LCQ & NCQM & FQL\\
\hline
\raisebox{-13pt}{\ }
 $0.43 \sim 0.48$  \raisebox{13pt}{\ } & $0.342 \pm 0.034$ & $0.37$
& 0.21 & $0.30 \sim 0.40$ & 0.16 \\
\hline\hline
\end{tabular}
\caption[]{
The predictions of QCD $S$ parameter in various models.
GL : experimental value by Gasser-Leutwyler,
VMD : vector and axial-vector dominance model,
LCQ : local chiral quark model,
NCQM : nonlocal constituent-quark model.
We also show the value calculated by free quark one-loop diagram
(FQL), $S = N_c/(6\pi)$.}\label{tab. S}
\end{center}
\end{table}

Now, let us compare our result with the experimental value
(GL)\cite{GL}
and other predictions from the vector meson dominance model
(VMD)\cite{Ecker-Gasser-Leutwyler-Pich-Derafael}, the local chiral
quark model (LCQ)\cite{Donoghue-Holstein}, and the nonlocal
constituent-quark model (NCQM)\cite{Holdom-Terning-Verbeek:PLB245,Holdom}.
We show their results in Table~\ref{tab. S}.\footnote[5]{
   They calculate $L_{10}(\mu)$ at the renormalization scale
   $\mu = m_\eta$ (LCQ, NCQM) or $\mu = m_\rho$ (VMD).
   Following Ref.~\cite{GL,Donoghue-Holstein},
   we convert their values of $L_{10}(\mu)$
   to $S$ by
   \[
   S = -16\pi \left[~L_{10}(\mu) +
   \frac{1}{192\pi^2}\left(\ln\frac{m_\pi^2}{\mu^2} +
   1\right)~\right]~.
   \]
}
The FQL or LCQ model gives a half of the experimental value (GL) of
$S$.
We should include higher order corrections as in
Figs.~\ref{fig. loop}(b) and \ref{fig. loop}(c).
The VMD model well reproduces the experimental value of $S$.
This implies that it is important to include the contribution from the
bound state, in other words, we should take into account the binding
force as in Fig.~\ref{fig. loop}(c).
On the other hand, in NCQM model the value of QCD $S$ parameter is
improved by the inclusion of the momentum dependent mass function as
in Fig.~\ref{fig. loop}(b).
The high energy behavior of this diagram is consistent with
the result by operator product expansion (OPE).\cite{OPE}
For these reasons we include these two corrections by the improved
ladder approximation.
Our value of QCD $S$ parameter is 30\% larger than the experimental
value.
For one thing further corrections beyond the improved ladder
approximation may be needed
(e.g., the decay widths of vector or axial-vector mesons are
not taken into account in our approximation);
for another the difference between our value and the experimental one
may be caused by the slight breaking of the chiral Ward-Takahashi
(WT) identity.
Reference~\cite{K-M} suggests that if we use the ladder approximation
completely consistent with WT identity we could make up the
difference.\footnote[5]{
There is 20\% difference in the value of $f_\pi$ between the
``consistent'' ladder approximation and Pagels-Stokar
formula\cite{K-M},
while Pagels-Stokar formula and the improved ladder approximation give
almost the same value of $f_\pi$\cite{A-K-M}.
[We note that the different BS kernels are used in two references.
For details see Refs.~\cite{A-K-M,K-M}.]
}

\subsection{ Pole Fitting }

We extract the mass $m_\rho$ and the decay constant
$f_\rho$ of $\rho$ meson by three-pole fitting:
\begin{equation}
\Pi(q_E^2) =
   \frac{f_\rho^2 m_\rho^2}{q_E^2+m_\rho^2} -
   \frac{f_{R_1}^2 m_{R_1}^2}{q_E^2+m_{R_1}^2} +
   \frac{f_{R_2}^2 m_{R_2}^2}{q_E^2+m_{R_2}^2}~, \label{eq. pole fit}
\end{equation}
from the two-point function
$\Pi(q_E^2) \equiv \Pi_{VV}(q_E^2) - \Pi_{AA}(q_E^2)$ in
space-like region ($0\leq q_E^2 \leq (1\mbox{GeV})^2$)
with $t_F = 0.5$.
The positive sign contribution in eq.~(\ref{eq. pole fit})
comes from the two-point function of the vector current
and the negative sign contribution from that of the axial-vector
current.
The last two terms represent the contributions from other mesons
heavier than $\rho$ meson.
The decay constant of $\rho$ meson in eq.~(\ref{eq. pole fit}) is
defined by
\begin{equation}
\Bigl\langle{0}\Bigr| V_\mu^a(0) \Bigl|{\rho^b(q,\epsilon)}\Bigr\rangle =
   \delta^{ab}\epsilon_\mu f_\rho m_\rho~.
\end{equation}

\begin{figure}[hbtp]
\begin{center}
\ \epsfbox{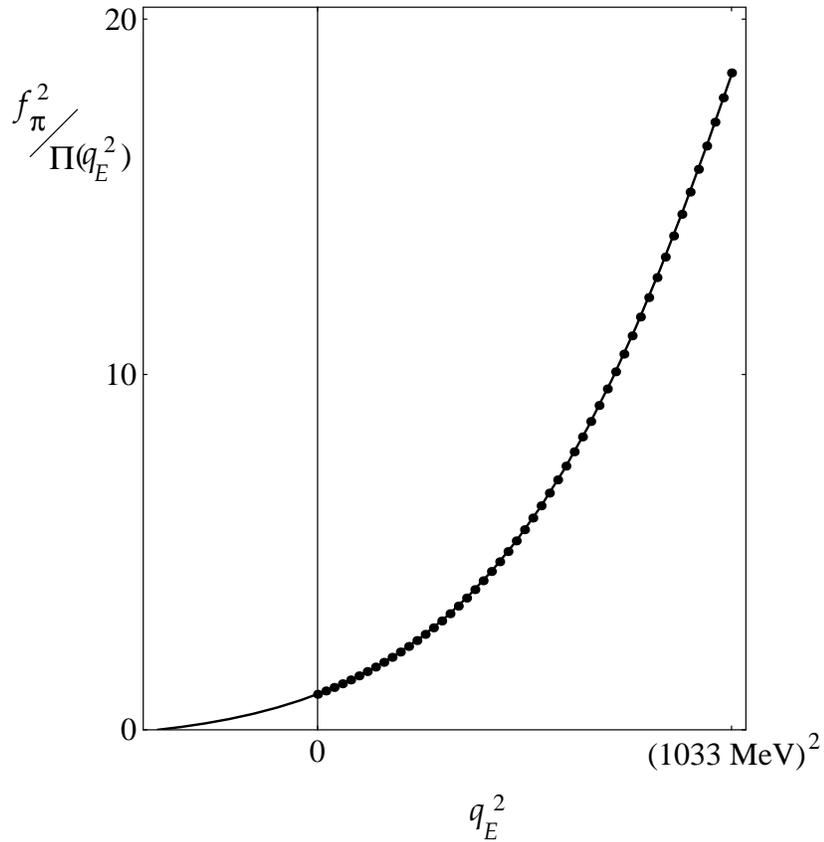}
\vspace{-7pt}
\caption[]{
Three pole fitting of our two-point function, $\Pi(q_E^2) =
\Pi_{VV}(q_E^2) - \Pi_{AA}(q_E^2)$.
The solid line is the best fitting curve, the dots denote the value of
our $\Pi(q_E^2)$.
}
\label{fig. V-A}
\end{center}
\end{figure}

We show our two-point function $\Pi(q_E^2)$ and the best
fitting curve in Fig.\ref{fig. V-A}.
The best fitted values are shown in Table~\ref{tab. rho}.
\begin{table}[hbtp]
\center{
\begin{tabular}[t]{|c||c|c|}
\hline
\raisebox{-8pt}{\ }\raisebox{13pt}{\ }
& Our value & Experiment~\cite{EXP} \\
\hline\hline
\raisebox{-8pt}{\ } $f_\rho$ [MeV] \raisebox{13pt}{\ }
& 133 & 144 $\pm$ 8 \\
\hline
\raisebox{-8pt}{\ } $m_\rho$ [MeV] \raisebox{13pt}{\ }
& 643 & 768 \\
\hline
\end{tabular}}
\caption[]{
The best fitted values of the mass and decay constant of $\rho$ meson.
We use our value of $\Lambda_{\rm QCD}$ ($ = 462$ [MeV]).
}\label{tab. rho}
\end{table}
These values should be compared with those in Ref.~\cite{A-K-M}.
Because we need no further regularization in calculating the ``$V-A$''
two-point function,
we do not have to introduce cutoff parameter as in
Ref.~\cite{A-K-M}.

We find that the sum of the pole residues
$f_\rho^2m_\rho^2 - f_{R_1}^2m_{R_1}^2 + f_{R_2}^2m_{R_2}^2$
vanishes, which implies that our $\Pi(q_E^2)$ behaves as $1/q_E^4$.
The result from the improved ladder approximation reproduces the high
energy behavior of the ``$V-A$'' two-point function required by that
from OPE in QCD.
This means that the spectral functions of our $\Pi(q_E^2)$ in
eq.~(\ref{eq. spectral repr.}) satisfies the second Weinberg sum rule:
\begin{equation}
\int_0^\infty ds \left[ \rho_V(s) - \rho_A(s) \right] = 0~.
\end{equation}
The masses and decay constants of the heavier mesons are unstable for
fitting;
we obtain several best fitted curves with different values for the
masses and decay constants of the heavier mesons,
although they satisfy the first and second Weinberg sum rules.
On the other hand, the lowest meson mass and decay constant are
very stable.

\begin{figure}[hbtp]
\begin{center}
\ \epsfbox{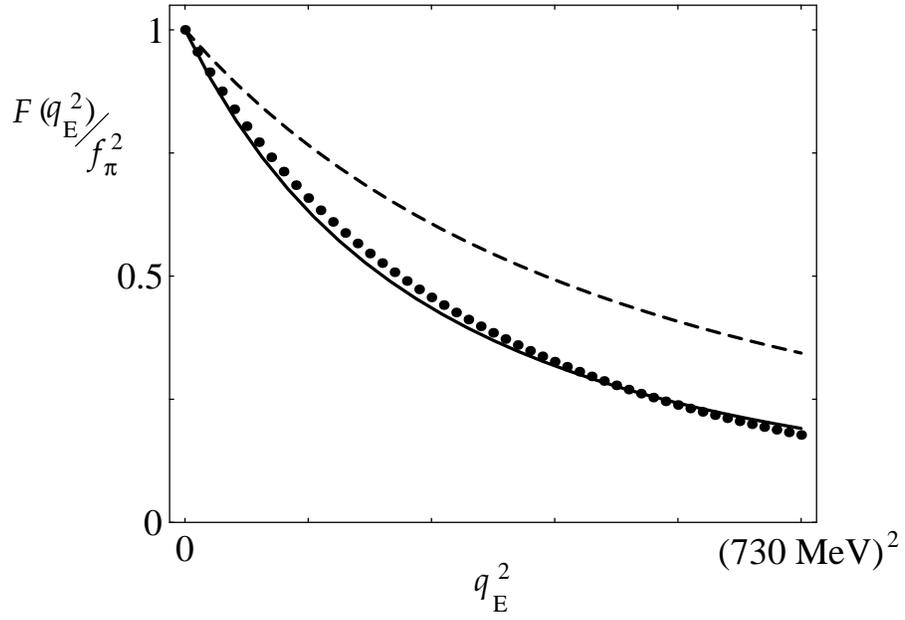}
\vspace{-7pt}
\caption[]{
The comparison with the $\rho$ and $a_1$ meson dominance model.
The solid line denotes $F(q_E^2)$ with
$(\coth\theta, m_\rho) = (1, 643\mbox{ [MeV]})$,
the broken line, $F(q_E^2)$ with $(f_\rho, m_\rho) = (144, 768)$
[MeV].
The dots denote the values of our $\Pi(q_E^2)$.
}\label{fig. vmd}
\end{center}
\end{figure}

We compare our two-point function with that from the $\rho$ and $a_1$
meson dominance model.
In this model the first and second Weinberg sum rules read
\begin{equation}
f_\rho^2 - f_{a_1}^2 = f_\pi^2~,\qquad
f_\rho^2 m_\rho^2 - f_{a_1}^2 m_{a_1}^2 = 0~.
\end{equation}
It is convenient to adopt the following parameterization:
\begin{equation}\begin{array}{ccc}
f_\rho = f_\pi \cosh\theta~, &
f_{a_1} = f_\pi \sinh\theta~.
\end{array}
\end{equation}
Then the resultant ``$V-A$'' two-point function $F(q_E^2)$ is
expressed by
\begin{equation}
F(q_E^2) = f_\pi^2 ~
   \frac{m_\rho^4 \coth^2\theta}
   {(q_E^2+m_\rho^2)(q_E^2+m_\rho^2\coth^2\theta)}~.
\end{equation}
We show the function $F(q_E^2)$ for two different choices of
$m_\rho$ and $\coth\theta$ in Fig.~\ref{fig. vmd} with our
$\Pi(q_E^2)$.
Taking $(\coth\theta, m_\rho) = (1,643\mbox{ [MeV]})$ makes $F(q_E^2)$
well agree with our $\Pi(q_E^2)$.
However, when we use the experimental value
$(f_\rho, m_\rho) = (144, 768)$[MeV], $F(q_E^2)$ does not match with
our $\Pi(q_E^2)$.

\subsection{ Walking Coupling Case }

We apply the method for inhomogeneous BS equation to the other
dynamical system than QCD, i.e., $SU(3)$ gauge theory with $N_D$
doublets of massless fermions.

The difference between the real QCD and the system which we
investigate in this section appears in the coefficient
of the $\beta$-function.
Namely, the coefficient factor $\alpha_0$ of the running coupling in
eq.~(\ref{eq. alpha}) is taken to be
\begin{equation}
\alpha_0 = \frac{4\pi}{\beta_0} = \frac{12\pi}{11N_c - 4N_D}~,
\end{equation}
where $\beta_0$ is the coefficient of one-loop $\beta$-function.
For the small value of $\beta_0$ the coupling runs very slowly, and
the system well simulates the walking technicolor
model\cite{H-YBM-AY-AKW}.

Using the same procedure as the previous one for real QCD,
we evaluate $S$ parameter and the pion decay constant with $N_D = 1,
\cdots, 6$.
The results are shown in Table~\ref{tab. ND dependence}.
\begin{table}[hbtp]
\center{
\begin{tabular}[t]{|c||c|c|c|c|c|c|}
\hline
$N_D$
   & 1     & 2     & 3     & 4     & 5     & 6     \\\hline\hline
$S/N_D$
   & 0.469 & 0.457 & 0.442 & 0.431 & 0.430 & 0.421 \\ \hline
$(f_\pi/\Lambda_{\rm QCD})^2 /N_D\times 100$
   & 3.87  & 4.27  & 4.87  & 6.05  & 9.32  & 24.7  \\ \hline
$\Lambda_{\rm QCD}$ [MeV]
   & 473  &  318   & 243   & 189   & 136   & 76.4  \\ \hline
\end{tabular}}
\caption[]{
The values of $S/N_D$ and $f_\pi/\Lambda_{\rm QCD}$ for the number of
doublets $N_D = 1, \cdots, 6$ with $t_F = 0.5$ fixed.
We also show the values of $\Lambda_{\rm QCD}$ by imposing
$f_\pi = 93$ MeV.
}\label{tab. ND dependence}
\end{table}
We find that the dominant support (cf. Fig.~\ref{fig. support}) lies
on the energy region higher for large $N_D$ than for small $N_D$.
The values of $\Lambda_{\rm QCD}$ become small for large
number of doublets.
However, we cannot see the clear dependence on $N_D$ of the values of
$S/N_D$.
The value slightly decreases as $N_D$ increased, only by 11\%.
Our result shows that $S$ parameter almost scales linearly with $N_D$,
which supports the naive estimation by Refs.~\cite{PT,Takeuchi}.

Finally we check the dependence on the infrared cutoff $t_F$ with
$N_D = 5$, and show the results in Table~\ref{tab. tF2 dependence}.
\begin{table}[hbtp]
\center{
\begin{tabular}[t]{|c||c|c|c|c|c|}
\hline
$t_F$
   & 0.3   & 0.5   & 0.7   & 0.9   & 1.1    \\\hline\hline
$S/N_D$
   & 0.450 & 0.430 & 0.432 & 0.436 & 0.446  \\ \hline
$(f_\pi/\Lambda_{\rm QCD})^2/N_D \times 100$
   & 8.82  & 9.32  & 9.78  & 10.2  & 10.4   \\ \hline
$\Lambda_{\rm QCD}$ [MeV]
   & 140   & 136   & 133   & 130   & 129    \\ \hline
\end{tabular}}
\caption[]{The dependence on the infrared cutoff $t_F$ for $N_D = 5$.}
\label{tab. tF2 dependence}
\end{table}
The dependence on the infrared cutoff $t_F$ of $\Lambda_{\rm QCD}$
should be compared with the result in the case of single-sextet quark
in Ref.~\cite{A-B-K-M-N}.
They calculate $f_\pi/\Lambda_{\rm QCD}$ using the homogeneous BS
equation of pion,
and find that the slowly running coupling gives stable results less
dependent on $t_F$.
Our results conform to theirs.


\vskip1cm
\centerline{\large\bf Acknowledgements}

We would like to thank T.~Kugo for useful discussions and comments.
M.H. is supported in part by the Grant-in-Aid for Scientific Research
(\#2208) from the Ministry of Education, Science and Culture.

\newcommand{\J}[4]{{#1} {\bf #2}, #4 (19#3)}
\newcommand{\MPL}{Mod.~Phys.~Lett.}
\newcommand{\NP}{Nucl.~Phys.}
\newcommand{\SJNP}{Sov.~J.~Nucl.~Phys.}
\newcommand{\PL}{Phys.~Lett.}
\newcommand{\PR}{Phys.~Rev.}
\newcommand{\PRL}{Phys.~Rev.~Lett.}
\newcommand{\PTP}{Prog.~Theor.~Phys.}
\newcommand{\AP}{Ann.~Phys. (N.Y.)}
\newcommand{\CMP}{Commun.~Math.~Phys.}


\begin{thebibliography}{99}

\bibitem{GL}
	J. Gasser and H. Leutwyler, \J{\AP}{158}{84}{142};
	\J{\NP}{B250}{85}{465}.

\bibitem{HT}
	B.~Holdom and J.~Terning, \J{\PL}{B247}{90}{88}.

\bibitem{PT}
	M. Peskin and T. Takeuchi, \J{\PRL}{65}{90}{964}.

\bibitem{AB}
	G.~Altarelli and R.~Barbieri, \J{\PL}{B253}{91}{161}.

\bibitem{Ecker-Gasser-Leutwyler-Pich-Derafael}
	J.F.~Donoghue, C.~Ramirez and G.~Valencia,
	\J{\PR}{D39}{89}{1947};
	G.~Ecker, J.~Gasser, H.~Leutwyler, A.~Pich and E.~de~Rafael,
	\J{\PL}{B233}{89}{425}.

\bibitem{Holdom-Terning-Verbeek:PLB245}
	B.~Holdom, J.~Terning and K.~Verbeek, \J{\PL}{B245}{90}{612}.

\bibitem{Holdom}
	B. Holdom, \J{\PR}{D45}{92}{2534}.

\bibitem{Donoghue-Holstein}
	J.F.~Donoghue and B. Holstein,
	\J{\PR}{D46}{92}{4076}.

\bibitem{Weinberg}
	S.~Weinberg, \J{\PRL}{18}{67}{507}.

\bibitem{A-B-K-M-N}
	K-I.~Aoki, M.~Bando, T.~Kugo and M.G.~Mitchard
	H.~Nakatani, \J{\PL}{B266}{91}{467}.

\bibitem{Higashijima}
	K.~Higashijima, \J{\PR}{D29}{84}{1228}.

\bibitem{Miransky}
	V.~Miransky, \J{\SJNP}{38}{84}{280}.

\bibitem{JM}
	P.~Jain and H.J.~Munczek, \J{\PR}{D44}{91}{1873}.

\bibitem{A-K-M}
	K-I.~Aoki, T.~Kugo and M.G.~Mitchard, \J{\PL}{B266}{91}{467}.

\bibitem{H-YBM-AY-AKW}
	B.~Holdom, \J{\PL}{B150}{85}{301};
	K.~Yamawaki, M.~Bando and K.~Matumoto, \J{\PRL}{56}{86}{1335};
	T.~Akiba and T.~Yanagida, \J{\PL}{B169}{86}{432};
	T.~Appelquist, D.~Karabali and L.C.R.~Wijewardhana,
	\J{\PRL}{57}{86}{957}.

\bibitem{AppelquistTriantaphyllou:PLB278}
	T.~Appelquist and G.~Triantaphyllou, \J{\PL}{B278}{92}{345}.

\bibitem{D-M-O}
	T.~Das, V.~Mathur and S.~Okubo, \J{\PRL}{19}{67}{859}.

\bibitem{Inami-Lim-Yamada}
	T.~Inami, C.S.~Lim and A.~Yamada, \J{\MPL}{7}{92}{2789}.

\bibitem{K-M-Y}
	T.~Kugo, M.G.~Mitchard and Y.~Yoshida, \J{\PTP}{91}{94}{521}.

\bibitem{GLUON}
	J.~Bijnens, C.~Bruno and E.~de~Rafael, \J{\NP}{B390}{93}{501}.

\bibitem{OPE}
	M.A.~Shifman, A.I.~Vainshtein and V.I.~Zakharov,
	\J{\NP}{B147}{79}{385}.

\bibitem{K-M}
	T.~Kugo and M.G.~Mitchard, \J{\PL}{B286}{92}{355}.

\bibitem{EXP}
	The experimental value of the $\rho$ meson mass is given in\\
	Particle Data Group: Review of Particle Properties,
	\J{\PR}{D45}{92}{};\\
	and the decay constant is given in\\
	J.~Gasser and H.~Leutwyler, \J{Phys.~Rep.}{87}{82}{77}.

\bibitem{Takeuchi}
	T.~Takeuchi, in {\it Proc. of International Workshop on
	Electroweak Symmetry Breaking}, Nov. 12-15, 1991, ed.
	W.A.~Bardeen, J.~Kodaira and T.~Muta (World Scientific Pub.
	Co., Singapore, 1991), p. 165.

\end{thebibliography}
\end{document}